\documentclass[%
 reprint,
 amsmath,amssymb,
 aps,
]{revtex4-2}

\usepackage[utf8]{inputenc}
\usepackage{graphicx}
\usepackage{amsmath}
\usepackage{comment}

\usepackage{graphicx}
\usepackage{subfig}

\begin{document}

\title{Cavities with Non-Spherical Mirrors for Enhanced Quantum Emitter-Cavity Photon Interaction}
\author{Denis V. Karpov}
\affiliation{Optoelectronics Research Centre, University of Southampton, Southampton SO17 1BJ, United Kingdom}

\author{Peter Horak}
\affiliation{Optoelectronics Research Centre, University of Southampton, Southampton SO17 1BJ, United Kingdom}
\email{peh@orc.soton.ac.uk}

\date{October 2021}

\begin{abstract}
We propose a procedure for the significant enhancement of the strong coupling rate between photons in an optical cavity and a single quantum emitter, such as an atom, quantum dot or trapped ion. We show that specially designed, non-spherical mirrors can lead to cavity eigenmodes that exhibit a large field enhancement at the center of the cavity while inducing significantly less beam divergence, and therefore smaller round trip losses and higher cooperativity, than can be achieved by operating a spherical-mirror cavity in the near-concentric regime. We verify our designs using mode matching theory and discuss their robustness relative to different kinds of manufacturing deviations.
\end{abstract}

\maketitle

\section{Introduction}

For various applications of modern quantum optics, both in experimental academic research and commercial quantum technology, strong coupling of a quantum emitter with an optical resonator is required, and simultaneously a long photon lifetime in this resonator is critically important. Some of the most promising systems to meet these requirements for practical applications are fiber-optic microcavities \cite{Pellizzari1995,Cirac1997,Kimble2008,Monroe2013}, ion beam etched dielectric resonators \cite{fib}, or micro-assembled structures \cite{revResonators}.

Strong coupling of a quantum emitter to an optical cavity for quantum technology applications can be achieved by tightly localising a single cavity photon, i.e., by making the cavity very small. However, for most realistic quantum information processing schemes optical access to the emitter from the side is required, e.g., for optical cooling \cite{cooling}, state preparation, and final state readout \cite{prep}. Moreover, channels for the delivery of atoms or ions into the cavity and the integration of trapping structures into the cavity may impose further constraints on the cavity length. In the case of ion-trap quantum computing \cite{comp}, the dielectric mirrors forming the cavity can also distort the radio-frequency fields required for trapping the ion due to their electric susceptibility and due to surface charges if they are too close to the trap electrodes \cite{Harlander2010,Podoliak2016}.

Overall, there is thus a need for optical cavities used in quantum information applications to combine a strong coupling rate with low losses, while at the same time also keeping the mirrors sufficiently apart. The goal of our work here is to present a new method to achieve these requirements by moving away from the paradigm of spherical mirrors to engineer optical cavity modes with better confinement properties than the standard Gaussian modes.  

Let us first review the principal parameters to characterize optical resonators for strong coupling to single particles. The coherent coupling between a two-level quantum emitter, such as a quantum dot, ion or cold atom, located at a coordinate $\textbf{r}$ in an cavity with an optical field mode $E(\textbf{r})$ is characterized by the single-photon strong coupling rate or coupling strength \cite{Vucovich}
\begin{equation}
g_0(\textbf{r})=\sqrt{\frac{3\lambda^2c\Gamma}{4\pi V_{\psi} }}\psi(\textbf{r}),\
\psi(\textbf{r}) = \frac{E(\textbf{r})}{|E(\textbf{r}_m)|}
\label{eq:coupl1}
\end{equation}
where $\Gamma$, $\lambda$, $L$ are the spontaneous dipole decay rate of the emitter $\Gamma = \frac{\omega^3 \mu^2}{3\pi\varepsilon_{0}\hbar c^3}$ (where $\omega$ is the transition angular frequency and $\mu$ is the electric dipole moment), its transition wavelength, and the cavity length, respectively. $V_{\psi}$ is the mode volume which we define as 
\begin{equation}
V_{\psi} = \frac{1} {|E(\textbf{r}_m)|^2 } \int  |E(\textbf{r})|^2 dV = \int |\psi(\textbf{r})|^2 dV
\label{eq:volume}
\end{equation}
where $\textbf{r}_m$ is the position of the maximum electric field in the cavity and the integration goes over the entire cavity volume. For a cavity mode with an emitter located at the maximum field intensity point $\textbf{r}_m$ the strong coupling rate is $g_0(\textbf{r}_m) = \sqrt{\frac{3\lambda^2c\Gamma}{4\pi V_{\psi}}}$.

To achieve strong emitter-cavity coupling, the coherent coupling rate $g_0$ between the emitter and the cavity mode must be larger than the strength of competing incoherent processes, i.e., larger than the dipole decay rate $\Gamma$ and the cavity loss rate $\kappa$. Therefore, the cooperativity parameter defined as
\begin{equation}
C_0 = \frac{g_0^2}{\kappa\Gamma}  = \frac{3\lambda^2c}{4\pi \kappa V_{\psi}} 
\label{eq:coop}
\end{equation}
must be larger than one.

We see from Eq.\ (\ref{eq:coop}) that the cooperativity can be increased by decreasing the mode volume. However, as discussed above, the cavity length $L$ must be sufficiently large to allow for optical side access and constraints imposed by atom/ion delivery and trapping. Therefore, if we write the mode volume as $V_\psi=L A_\psi$, reducing the mode volume can only be achieved by reducing the mode cross section $A_\psi$ (at the position of the particle). The common approach for this is to operate the cavity in a near-concentric regime \cite{yariv} where, in paraxial approximation, the mode cross section can be made arbitrarily small. On the other hand, this cavity configuration leads to excessive mode divergence and therefore to increased cavity losses $\kappa$ because large fractions of the mode fields miss the resonator mirrors, which eventually reduces the cooperativity again according to Eq.\ (\ref{eq:coop}). Furthermore, concentricity leads to cavity instability \cite{yariv} and thus makes cavity performance sensitive to smallest alignment errors.

Here we propose another method to increase the field amplitude in the center of the cavity $\psi(0)$, and thus to enhance cooperativity: by modifying the shape of the cavity mirrors we generate optimized interference patterns in the mode field. We explore modulated spherical mirror profiles that give rise to highly localized cavity modes while at the same time providing cavity loss rates far below those obtained with near-concentric cavities with comparable field localization for a wide range of parameters.

This paper is organized as follows. First, in Sec.\ \ref{sec:statement}, we describe our theoretical model and compare the enhancement of the strong coupling rate, mode losses, and cooperativity of target mode superpositions compared to near-concentric cavities. In Sec.\ \ref{sec:results} we discuss how to design non-spherical mirrors to generate cavity eigenmodes that consist of such mode superpositions and we verify those designs with a-priory simulations. We also investigate the stability of our designs against fabrication errors and briefly discuss potential fabrication methods. Finally we summarize and conclude in Sec.\ \ref{sec:conclusion}.


\section{Cavity Field Enhancement by Mode Superpositions}\label{sec:statement}


\subsection{Mode Matching Theory}\label{sec:formalism}

We start our investigation by introducing the formalism used later in the paper to calculate the eigenmodes of an optical cavity with arbitrarily shaped mirrors, following the approach of Refs.\ \cite{Kleckner2010, Nina}.

We write the optical mode fields $\Psi(\textbf{r})$ as a superposition of a given set of basis modes $\{\psi_n(\textbf{r})\}$, 
\begin{equation}
\Psi (\textbf{r})=\sum_{n}T_{n}\psi_n(\textbf{r}).
\label{eq:mode}
\end{equation}
We will consider only linearly polarized modes and thus the optical field can be described by a scalar field $\Psi(\textbf{r})$. For simplicity we also assume cylindrical symmetry of the cavity and thus restrict our model to radially symmetric basis modes. For mirror shapes that are modifications of standard spherical profiles, it is thus convenient to use the well-known Laguerre-Gaussian (LG) modes of symmetric, spherical-mirror cavities as the basis modes in our model. The basis mode functions are thus defined as
\begin{eqnarray}
\psi_{n}^{\pm}(\rho, \zeta)&=&\sqrt{\frac{2}{\pi}}L_n(2\rho^2) \times\nonumber \\
& & 
\exp{\{-\rho^2\pm i(-\zeta\rho^2+(2n+1)\tan^{-1}\zeta)}\}
\label{eq:basis}
\end{eqnarray}
where $\rho=r/w(\zeta)$ and $\zeta=z/z_0$ are the dimensionless radial and axial coordinates, $w(\zeta)=w_0\sqrt{1+\zeta^2}$ is the basis mode radius, $w_0$ is the mode waist of the fundamental mode, $z_0 =k w_0^2/2$ is the Rayleigh length, $k=2\pi/\lambda$ is the wave number, $L_n$ are Laguerre polynomials and $n$ is the mode order. The index $\pm$ marks the propagation direction.

The change of an optical field undergoing one round trip through the cavity can be represented by a mode-mixing operator $M$. Then finding the cavity eigenmodes reduces to solving the eigenproblem~\cite{Benedikter2015, Kleckner2010}
\begin{equation}
M\Psi = \gamma\Psi
\label{eq:ep}
\end{equation}
where the eigenvectors are the cavity modes expressed as superpositions of the basis LG modes and the eigenvalues $\gamma$ define the field amplitude changes of the corresponding modes acquired per round trip. The fractional loss per round trip $D$ for the mode $\Psi$ is given by
\begin{equation}
D=1-|\gamma|^2.
\label{eq:loss_per_roundtrip}
\end{equation}
The mode-mixing matrix of the cavity is given by
\begin{equation}
M=\exp{\{2ikL}\} A B,
\label{eq:mm1}
\end{equation}
where the exponential represents the longitudinal phase shift acquired over one cavity round trip, i.e., by propagation through twice the cavity length. The matrices 
$A$ and $B$ represent the left and right-side mirrors of the cavity positioned at coordinates $z=-L/2$ and $+L/2$, respectively, and are given by the mode overlap integrals taken over the finite extent of the mirrors, 
\begin{eqnarray}
A_{n,m} &=& \int_0^{\rho_a} \psi_n^+ \psi_m^{- *}\exp{\{ -2ik \Delta(\rho)\}}\, 2\pi\rho \, d\rho \Bigr|_{\substack{\zeta=-\zeta_m}},
\label{eq:mm2} \\
B_{n,m} &=& \int_0^{\rho_a} \psi_n^- \psi_m^{+ *}\exp{\{ +2ik \Delta(\rho)\}}\, 2\pi\rho \, d\rho \Bigr|_{\substack{\zeta=\zeta_m}},
\label{eq:mm3}
\end{eqnarray}
where $\rho_a=R_a/w(\zeta_m)$ with the mirror radius $R_a$ and the axial mirror position $\zeta_m= \frac{L}{2 z_0}$. $\Delta(r)$ describes the deviation of the mirror profile from a plane surface.

The calculation of the cavity modes consists of the following steps. As we are interested in cavity mirrors that are close in shape to a spherical profile, we first define a mirror radius of curvature $R$ and a cavity length $L$. We then calculate the waist $w_0$ of the fundamental mode of this spherical-mirror cavity and use this to define the fundamental and higher order modes $\psi_{n}^{\pm}$, Eq.\ (\ref{eq:basis}). The overlap integrals, Eqs.\ (\ref{eq:mm2})-(\ref{eq:mm3}) are then calculated numerically in Matlab for the modified mirror shapes discussed later in Section \ref{sec:results}. Finally, the full matrix $M$ and its eigenvalues and eigenvectors are calculated. We normalize the eigenvectors $\Psi$ to have the same mode volume $V_{\Psi}$ as the LG modes $V_{\psi}$ (note that all LG modes have the same mode volume),
\begin{equation}
V_{\Psi}=V_{\psi} = \frac{\pi}{4}w_0^2 L
\label{eq:volumes}
\end{equation}
where the last equality is evaluated from the mode definitions (\ref{eq:basis}). 
In this case the coherent coupling rate for $\Psi$ has the same form as Eq.\  (\ref{eq:coupl1}), i.e.,
\begin{equation}
g(\textbf{r})=\sqrt{\frac{3\lambda^2c\Gamma}{4\pi V_{\psi} }}\Psi(\textbf{r}).
\label{eq:coupl2}
\end{equation}
The eigenvectors of matrix $M$ contain the mode composition coefficients $T_{n}$ and thus the field profile of any cavity eigenmode at any point in the cavity can be found using Eq.~(\ref{eq:mode}). The corresponding cavity loss is obtained from the eigenvalue as per Eq.~(\ref{eq:loss_per_roundtrip}).


\subsection{Comparison of Mode Superpositions and Concentric Cavities}
\label{sec:advantages}

\begin{figure}[tb]
  \centering
  \includegraphics[height=5cm]{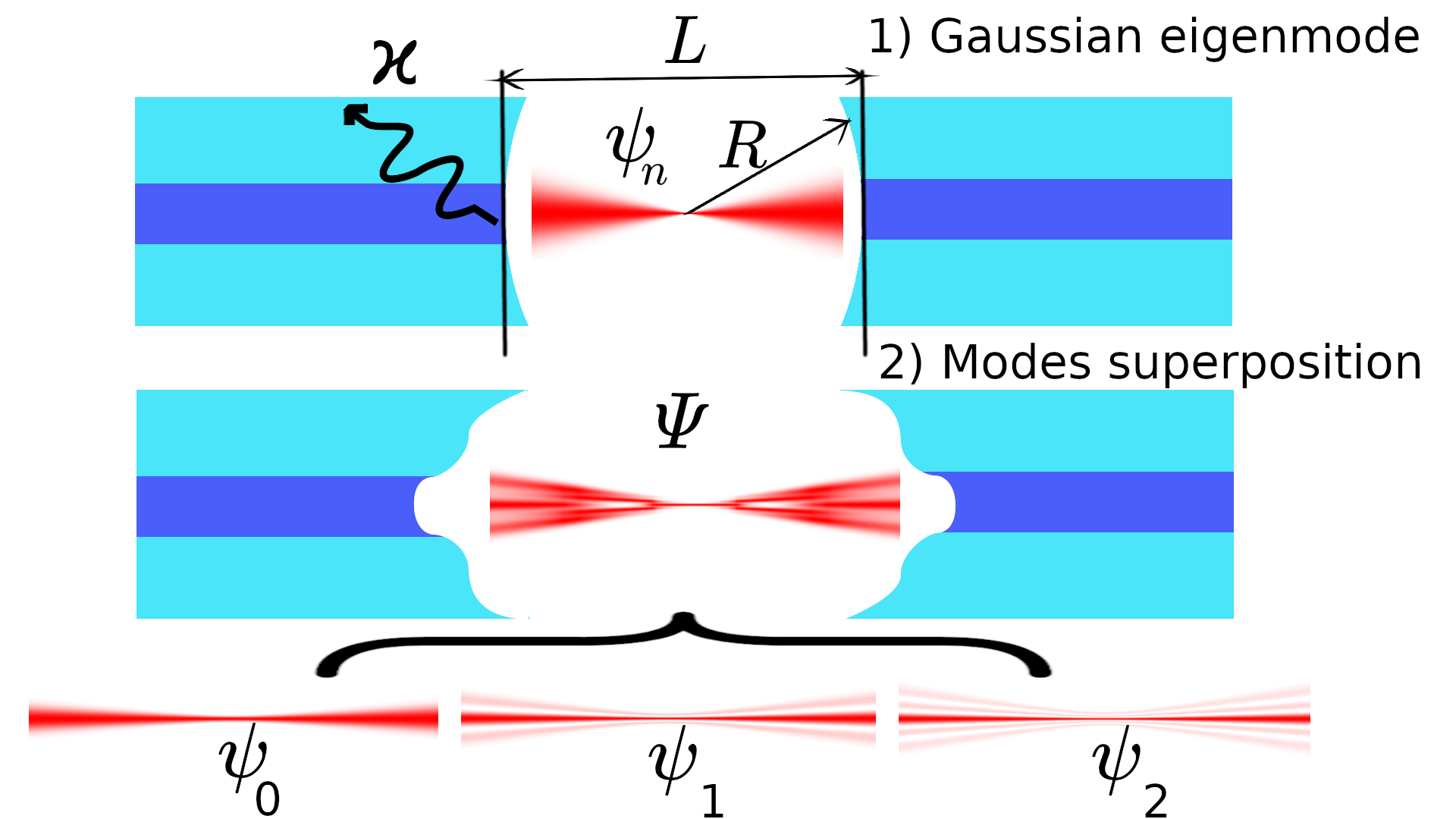}
  \caption{Schematic representation of 1) a spherical-mirror cavity and 2) a cavity with modified mirror shape obtained by our formalism. In case 1) the fundamental mode of the cavity is a fundamental Gaussian mode, in case 2) it is a superposition of the fundamental and of higher order modes. In both cases the cavities are assumed to be coupled to optical fibres at the mirror outputs.}
  \label{fig:cavity}
\end{figure}

We are interested in comparing two methods to increase the intensity of the field at the center of the cavity as discussed in the following

1) The ``traditional'' approach is to operate the cavity in the near concentric regime and use the fundamental Gaussian mode of the cavity, see top of Fig.~\ref{fig:cavity}. In this case the strong coupling rate scales as $g_0 \sim 1/w_0$ where $w_0$ is the waist at the center. The mode spot size on the mirrors (in the far field where $L/2$ is much larger than the Rayleigh length) scales with $1/w_0$ \cite{yariv}. So the spot size on the mirrors increases proportional to $g_0$. Once the spot size reaches the finite radius of the cavity mirrors, the majority of the light field misses the mirror, leading to clipping losses that increase exponentially with the spot size and thus with $g_0$. Thus, setting the cavity close to concentric maximizes the cavity field in the center, but rapidly increases the clipping loss and therefore strongly reduces the cooperativity.

2) The alternative method we are investigating here is using a superposition of higher order LG modes, see bottom of Fig.~\ref{fig:cavity}. Let us assume that we can shape the mirrors in such a way that the cavity supports an eigenmode that is an equal superposition of the first $N$ LG modes (we will discuss in Section \ref{sec:results} how this can be achieved). Hence, the eigenmode takes the form
\begin{equation}
\Psi (\rho, \zeta)=
\frac{1}{\sqrt{N}}\sum_{n=0}^{N-1} \psi_n^+(\rho, \zeta)
\label{eq:anzats}
\end{equation}
for the right-propagating field, and analogously for the left-propagating field. Note that the mode is normalized to maintain the mode volume of a single LG mode. For this mode, the strong coupling rate $g(0)$ at the center, Eq.\ (\ref{eq:coupl2}), scales with $\sqrt{N}$ because for all Laguerre polynomials we have $L_n(0)=1$ in the mode ansatz in Eq.\ (\ref{eq:anzats}).
On the other hand, the spot size of n-th LG mode on the mirror scales as $\sqrt{(2n+1)}$ \cite{Phillips1983}. Thus, the spot size of the highest order mode used in the superposition (\ref{eq:anzats}) scales as $\sqrt{N}$ and therefore proportional to $g(0)$. This is the same scaling as for the concentric approach discussed above, however, in the case considered here the $N$-th LG mode only carries $1/N$ of the power. It is therefore expected that for the equal mode superposition (\ref{eq:anzats}) the clipping losses due to the increasing mode spot size on the mirror with increasing strong coupling rate are smaller than for the near-concentric case.

\begin{figure}[tb]
  \centering
  \includegraphics[height=6cm]{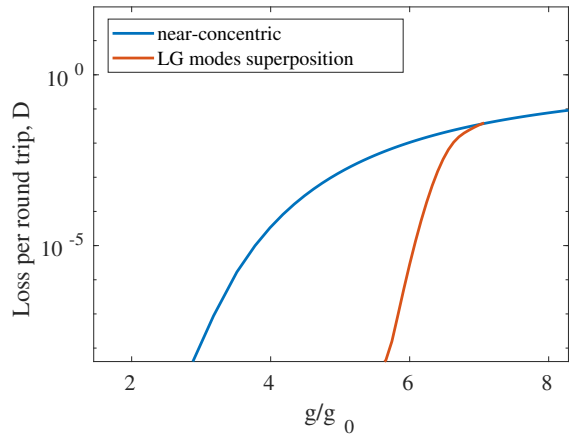}
  \caption{Clipping loss as a function of coupling rate (local electric field) enhancement factor. For the near-concentric cavity approach (blue curve) $g$ is enhanced by reducing the mirror radius of curvature towards $L/2$. For the mode superposition approach (red curve) $g$ is enhanced by increasing the number of modes in the superposition. Enhancement of $g$ is relative to the value $g_0$ of a reference cavity with $L = 500\,\mu$m, $R = 255\,\mu$m, $R_a= 100~ \mu$m, $\lambda=0.866~ \mu$m. }
  \label{fig:statement}
\end{figure}

We show a numerical example of the scaling of clipping losses with increased strong coupling rate in Fig.\ \ref{fig:statement}, where the clipping losses are calculated numerically from evaluating the mode overlap with the mirrors with finite radius $R_a$. Here we start with a reference cavity with cavity length $L=500\,\mu$m, mirror radius of curvature $R=255\,\mu$m, mirror radius $R_a=100 ~\mu$m, operating at wavelength $\lambda= 0.866~\mu$m. The strong coupling rate of this reference cavity at the center is denoted $g_0$ and the clipping losses are below the numerical accuracy in our simulations of $10^{-9}$ per round trip. For the near-concentric cavity approach, blue curve in Fig.\ \ref{fig:statement}, we then scan the radius of curvature towards $L/2=250\,\mu$m. This increases the strong coupling rate $g$ at the center but simultaneously increases the cavity round trip losses. For the mode superposition approach, red curve in Fig.\ \ref{fig:statement}, we increase the number of constituent modes $N$ in Eq.\ (\ref{eq:anzats}) from 1 to 50. As discussed above, this increases the strong coupling rate $g$ proportional to $\sqrt{N}$. We can clearly see that the corresponding increase in losses is significantly below the near-concentric approach - over a wide range of parameters comparable strong coupling amplification factors can be achieved with orders of magnitude lower losses. 


\subsection{Coupling Rate Enhancement}\label{sec:gsec}

In Fig.\ \ref{fig:statement} we considered strong coupling rate enhancement relative to a specific reference cavity of given mirror radius of curvature. We now investigate how the enhancement depends on radius of curvature in more detail.

\begin{figure}[tb]
  {(a) \hspace{5cm}}\\
  \includegraphics[height=6cm]{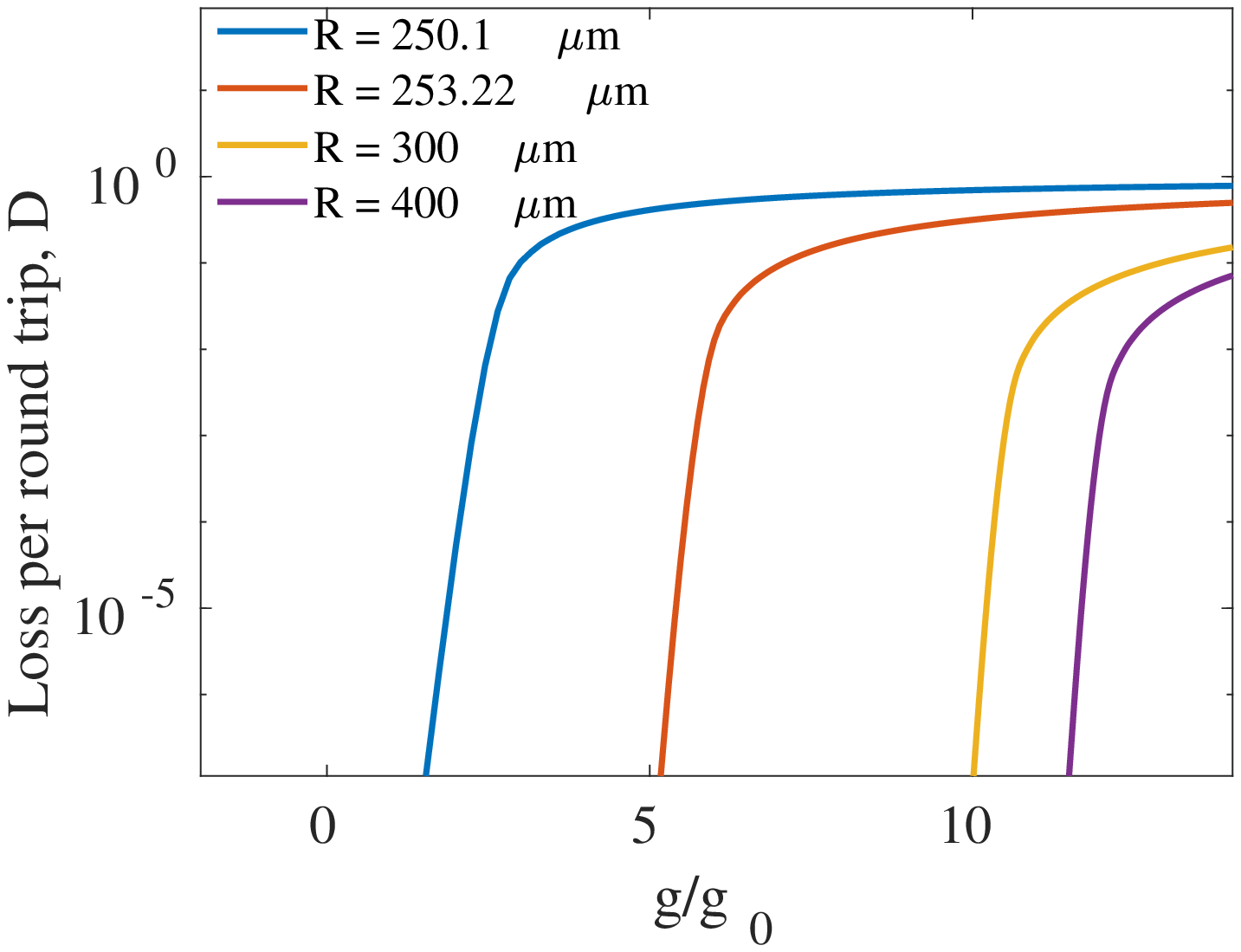}\\
  {(b) \hspace{0cm}}\\
  \includegraphics[height=6cm]{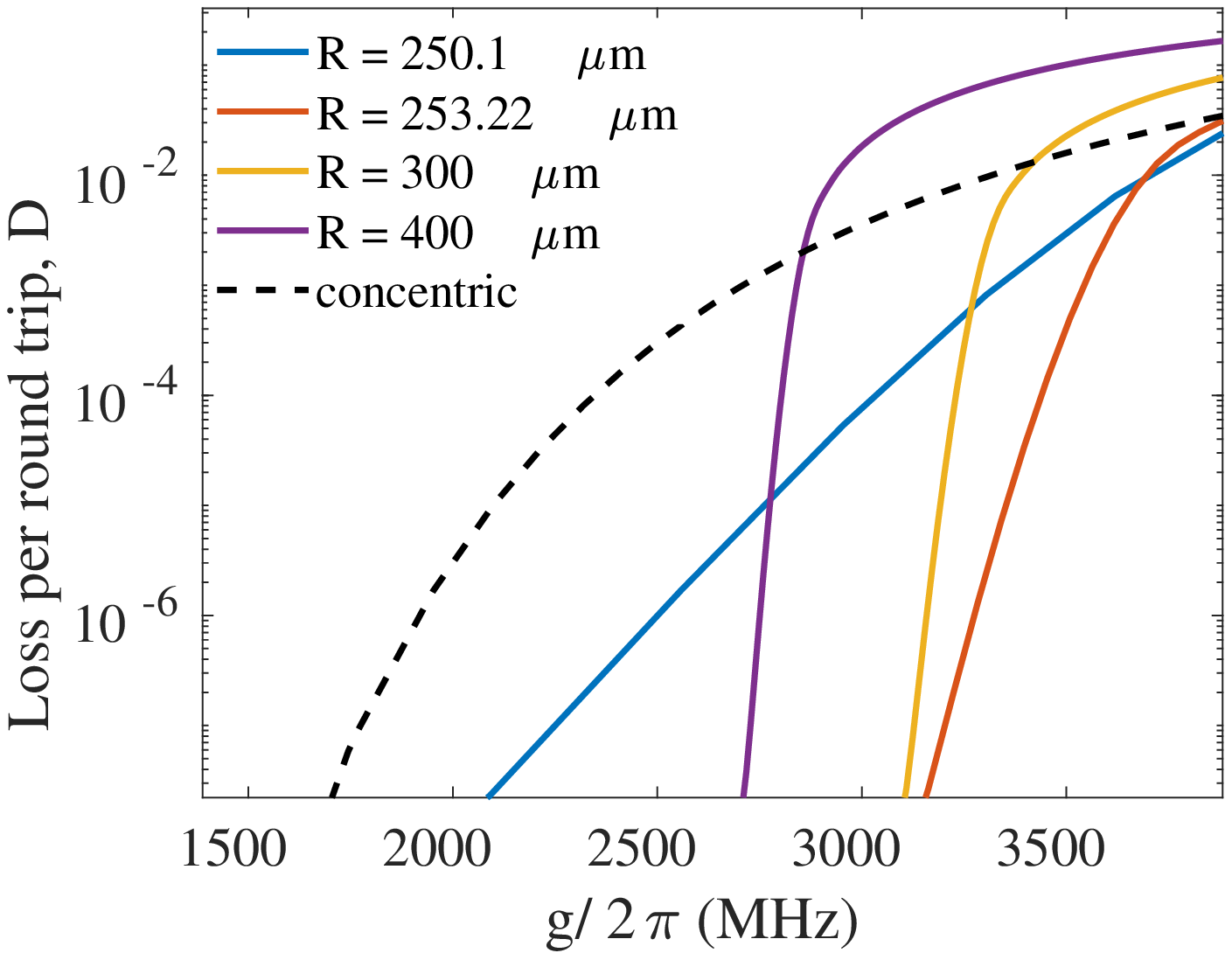}
  \caption{(a) Clipping losses as a function of enhancement of strong coupling rate for different radius of curvature, $L = 500~\mu m$.
  (b) Clipping losses as a function of strong coupling rate of Ca$^+$ ion for different radius of curvature, $L = 500~\mu m, \Gamma_{Ca^+} = 22$ MHz/$2\pi$.}
  \label{fig:radius_new}
\end{figure}

Figure \ref{fig:radius_new}(a) shows the enhancement of the strong coupling rate $g/g_0$ and the corresponding increase in clipping losses of the mode superposition approach for different reference cavities, i.e., cavities with different mirror radius of curvature. As before, the enhancement in $g/g_0$ is achieved by increasing the number $N$ of modes in the mode superposition. For near-concentric reference cavities, e.g. for $R=250.1\,\mu$m, only a modest enhancement in $g$ can be realized before the round trip losses increase significantly. In this limit, already the fundamental mode has a large spot size on the mirrors and thus adding more higher order modes to the superposition will quickly lead to excessive losses. As we move away from the concentric regime by increasing the mirror radius of curvature, the fundamental mode spot size on the mirrors decreases and thus more higher order modes can be added before losses increase substantially.

Note, however, that for increasing radius of curvature $R$ the strong coupling rate $g_0$ of the reference cavity decreases and thus every curve in Figure \ref{fig:radius_new}(a) is relative to a different value $g_0$. In Figure \ref{fig:radius_new}(b) we therefore plot the data versus $g$ in real units, where we use the parameters for the $866$-nm transition in Ca$^+$, a species frequently used in ion trap quantum information experiments \cite{CaIon}.
As mentioned above, for a reference cavity very close to the concentric limit the benefit of using higher-order mode superpositions is reduced. As the radius of curvature is increased, larger strong coupling rates become accessible at lower losses. Approaching a more confocal arrangement we can significantly enhance $g$ and keep losses under control, which corresponds to the region below the curve for the concentric cavity (black dashed line) in Figure \ref{fig:radius_new}(b). However, we observe that for too large radii of curvature (300 or 400 $\mu$m in the figure) the achievable enhancement factor reduces again.

Figure \ref{fig:radius_new}(b) shows clearly that there are parameter regions of cavities with larger strong coupling rate and lower loss that can be achieved with the mode superposition approach but are not accessible by near-concentric resonators. Finally, we note, however, that in order to achieve large strong coupling rate enhancement for cavities with relatively large mirror radius curvature, a very large number of higher order modes need to be included in the mode superposition. This may be difficult to realize, as discussed in Section \ref{robust}.


\subsection{Cooperativity Enhancement}\label{sec:csec}

Next we consider the enhancement of the cooperativity using the mode superposition approach for a more realistic situation which takes into account intrinsic cavity round trip losses because of light absorption or partial transmission through the mirrors, which typically are in the range of $D_{mir} = 10^{-5}-10^{-4}$. 
In this case the total loss rate $\kappa$ relates to both the loss of the mirrors $D_{mir}$ and the clipping loss per round trip $D(N)$ which depends on the number of LG modes in representation (\ref{eq:anzats}),
\begin{equation}
\kappa = \frac{c}{2 L} (D(N) + D_{mir}).
\label{eq:kappa}
\end{equation}
Using Eqs.\ (\ref{eq:volumes})-(\ref{eq:kappa}), the cooperativity $C$, Eq.\ (\ref{eq:coop}), then becomes 
\begin{equation}
C = \frac{6}{\pi^2} \left( \frac{\lambda}{w_0}\right)^2 \frac{N}{D(N) + D_{mir}}.
\label{eq:coop2}
\end{equation}
We now denote the cooperativity of the fundamental Gaussian mode as $C_0$ and assume it has negligible clipping losses, $D(N=1)=0$. The enhancement of cooperativity by an $N$-mode superposition is finally expressed as 
\begin{equation}
\frac{C}{C_0} =  \frac{N}{\frac{D(N)}{D_{mir}} + 1}.
\label{eq:coop3}
\end{equation}

\begin{figure}[tb]
  \centering
  \includegraphics[height=6cm]{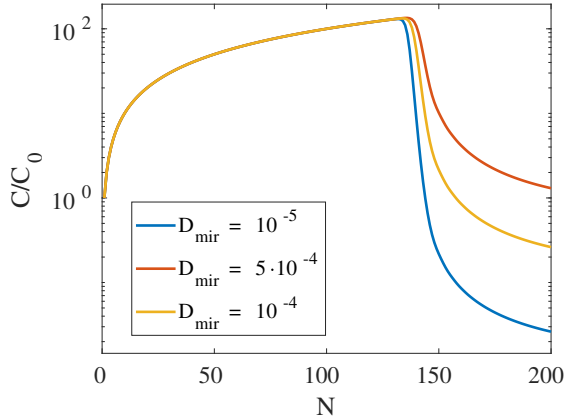}
  \caption{Cooperativity enhancement as a function of the number of modes $N$ in the superposition (\ref{eq:anzats}) for different intrinsic mirror losses $D_{mir}$. Other parameters are L$ = 500~\mu$m, $R = 400~\mu$m, $\lambda= 0.866~\mu$m.}
  \label{fig:coop}
\end{figure}

We show this cooperativity enhancement for different values of $D_{mir}$ in Fig.\ \ref{fig:coop}. As long as the clipping losses are smaller than the intrinsic losses, $D<D_{mir}$, the cooperativity increases linearly with $N$ and is independent of $D_{mir}$. Thus, a superposition of 100 modes would ideally enhance the cooperativity by two orders of magnitude. In practice, the region of $N$ between 0 and 10 is the most interesting because it provides a significant cooperativity enhancement and at the same time does not require too many modes, which would pose excessive constraints on the fabrication accuracy of the mirrors. 

The cooperativity starts to drop sharply with increasing $N$ when the clipping losses exceed $D_{mir}$. This happens when the $N$-th order mode spot size on the mirror \cite{Phillips1983},
\begin{equation}
    w_N(L/2) = w_0\sqrt{2N+1}\sqrt{1+[L/(2z_0)]^2},
\end{equation}
becomes comparable to the mirror radius $R_a$. This allows us to derive an estimate for a maximum useful number of modes in the superposition, 
\begin{equation}
N_\mathrm{cut off} \approx 2\left(\frac{\pi w_0 R_a}{\lambda L}\right)^2.
\end{equation}
This value of $N_\mathrm{cut off}$ thus marks the position of the sharp cutoff observed in Fig.~\ref{fig:coop}.


\section{Design of Mirror Profiles}\label{sec:results}

In the previous section we showed that certain superpositions of LG modes exhibit a large field enhancement in the cavity center while maintaining low cavity loss rates. We will now discuss the shape of the cavity mirrors that are required to generate such mode superpositions as eigenmodes of an optical cavity.


\subsection{Mode Field Phase Fronts}
\label{sec:geom}

\begin{figure}[tb]
  \centering
   {(a) \hspace{5cm}}\\
  \includegraphics[height=4cm]{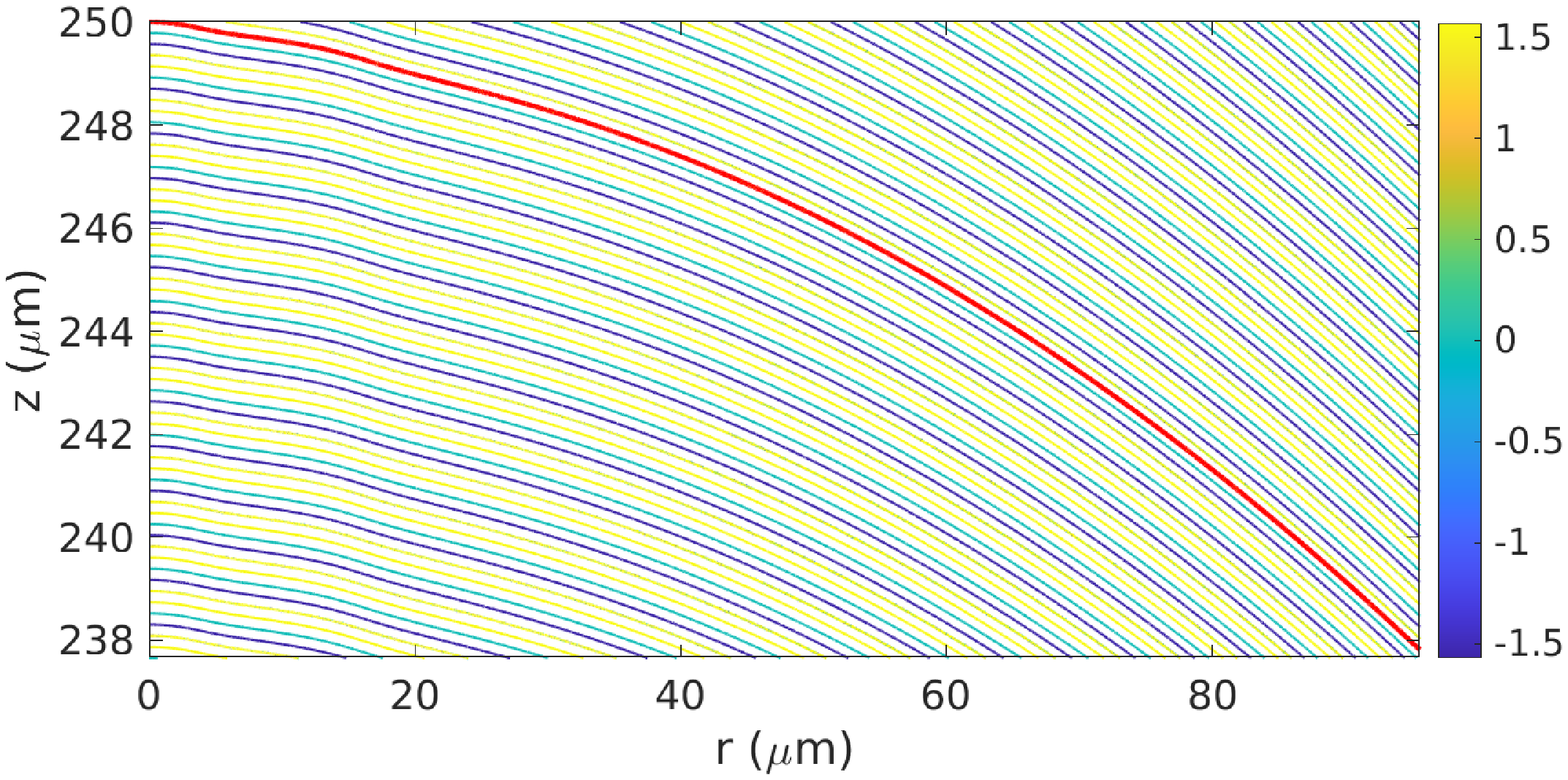}\\
    {(b) \hspace{0cm}}\\
  \includegraphics[height=3cm]{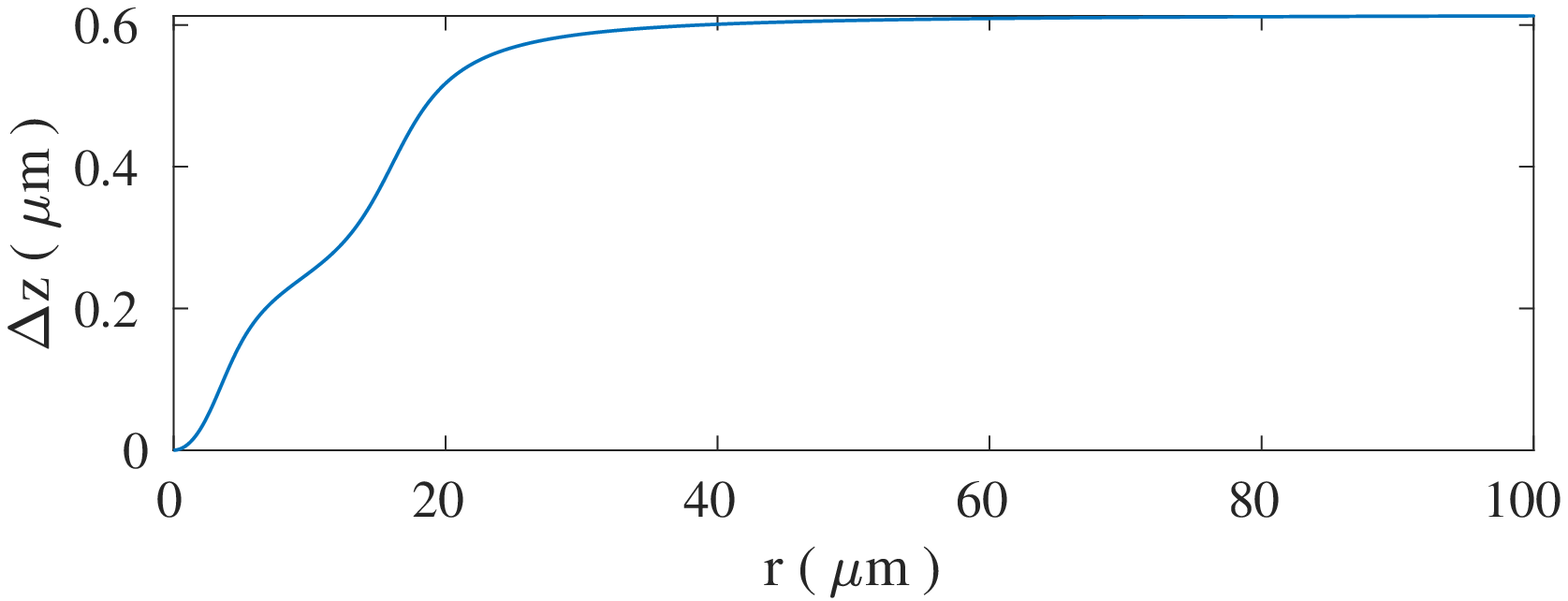}
  \caption{Phase of the target field (superposition of $N=3$ LG modes) near one of the cavity mirrors. The red curve marks a contour of constant phase, which would be an appropriate mirror profile to generate this field as a cavity eigenmode. (b) Shape of the required mirror profile as a deviation from a spherical profile. Parameters are L$ = 500~\mu$m, $R = 400~\mu$m, $\lambda= 0.866~\mu$m.}
\label{fig:phase}
\end{figure}

The principle of designing a mirror that will generate a desired electromagnetic field as an eigenmode of the cavity is as follows. We first calculate the desired field and analyze its wave fronts in the area of one mirror; an example is shown in Fig.\ \ref{fig:phase}(a). The required mirror shape then needs to follow a surface of constant phase, as presented by the red curve in the figure. 

For a field composed of a low number of LG modes, $N=3$ in Fig.\ \ref{fig:phase}, this mirror shape still mainly follows a spherical profile. It is thus instructive to plot the mirror shape as the deviation from the spherical profile, shown in Fig.\ \ref{fig:phase}(b). We see that the mirror can effectively be fabricated by manufacturing a ``hole'' of a certain shape and a maximum depth of the order of one micron into the center of a spherical mirror. We will further discuss possible routes to fabrication and the required manufacturing precision in Sec.\ \ref{fab} and Sec.\ \ref{robust}, respectively.

Note that all calculations in this paper are based on the paraxial approximation \cite{Lax1975}, and therefore are only valid for fields with low divergence angles. Within this approximation a spherical mirror is equivalent to a parabolic profile. Moreoever, the mode mixing matrices $A$ and $B$, Eqs.\ (\ref{eq:mm2})-(\ref{eq:mm3}), are calculated in the limit of thin mirrors, i.e., by applying a phase shift on a plane of constant position $z=L/2$. Numerically, instead of extracting a contour of constant phase as described above, we calculate the phase of the target field along $z=L/2$ and use this phase for the terms $\exp\{\pm 2ik\Delta(\rho) \}$ in Eqs.\ (\ref{eq:mm2})-(\ref{eq:mm3}). Within the paraxial and thin-mirror approximations the two approaches are equivalent.


\subsection{Verification of Mirror Designs}
\label{sec:verif}

Having extracted the shape of the mirror from the wave front pattern near $z=L/2$, as described above, we now use it to verify our mirror design. Using the formalism presented in Sec.\ \ref{sec:formalism} we calculate the eigenmodes of the cavity formed by two identical mirrors with the extracted shape and analyze the clipping losses $D$ of the eigenmodes, Eq.~(\ref{eq:loss_per_roundtrip}), and their coefficients $|T_i|^2$ which describe the weight of each LG mode $i$ in each cavity mode.

\begin{figure}[tb]
  {(a) }\\
  \includegraphics[height=5cm]{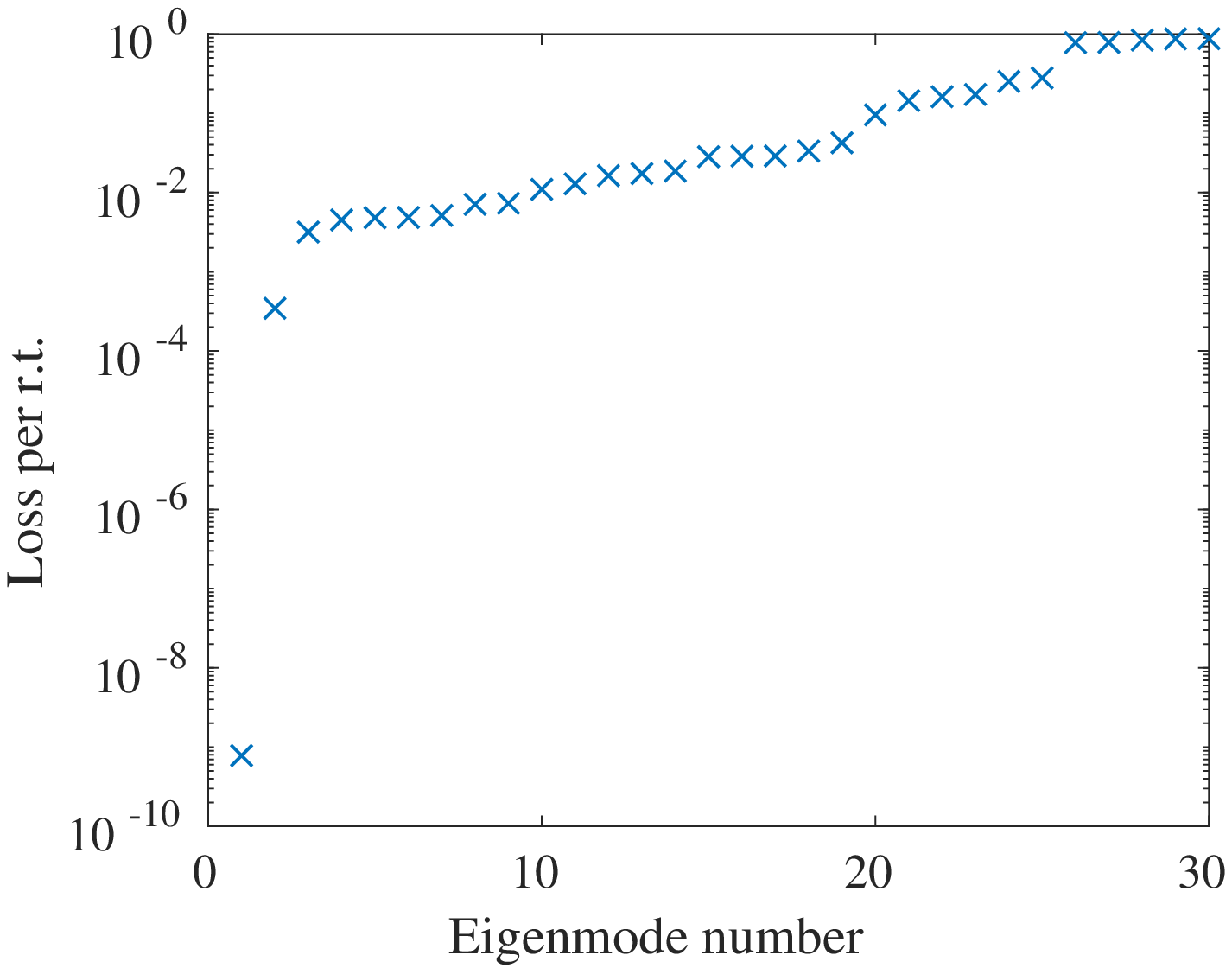}\\
   {(b) }\\
  \includegraphics[height=5cm]{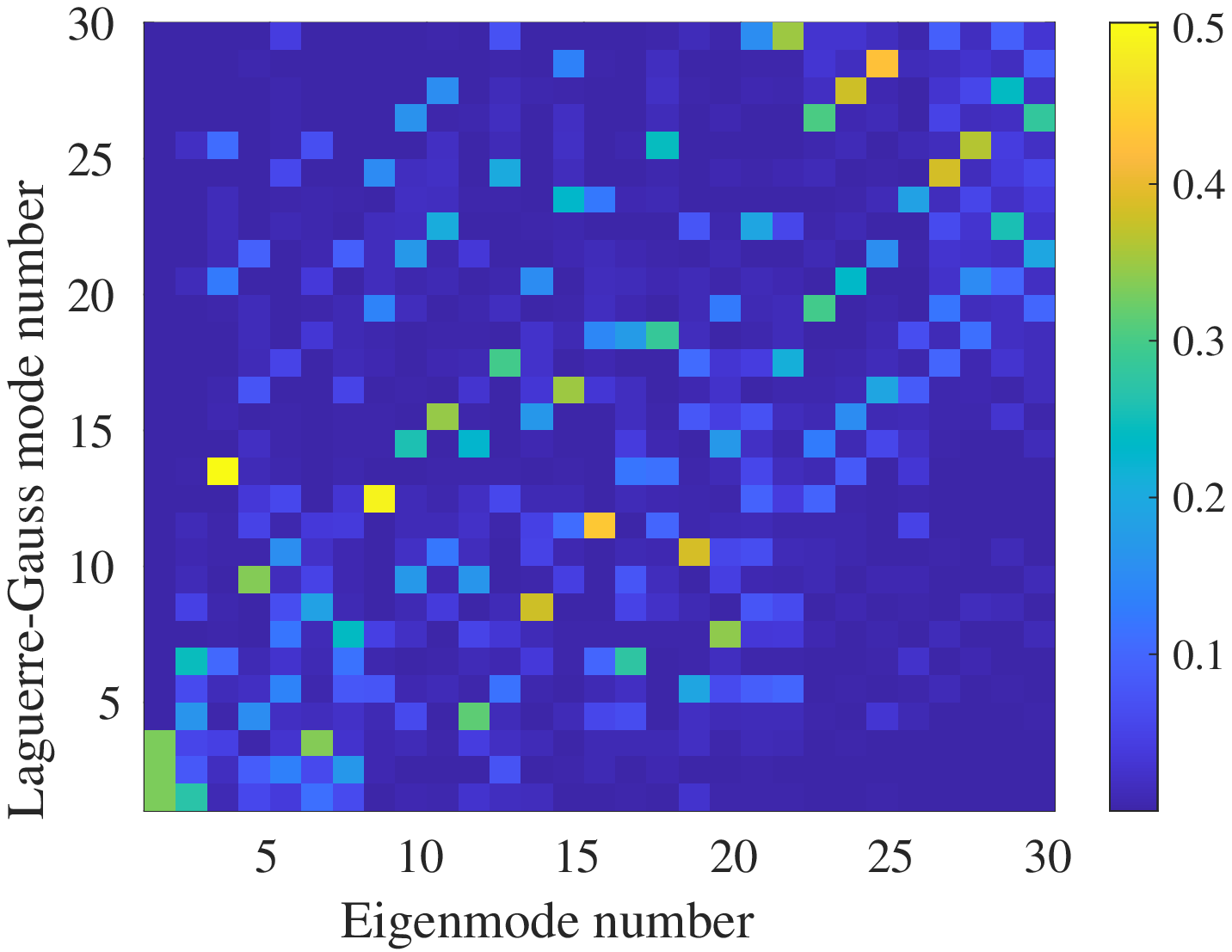}
  \caption{Eigenmodes of a cavity formed by mirrors designed in Fig.\ \ref{fig:phase} using mode matching theory. (a) Clipping loss per round trip $D$ for different eigenmodes sorted by increasing loss. (b) Corresponding coefficients $|T_i|^2$. The lowest loss eigenmode has weights 1/3 for the lowest 3 LGs, confirming the target mode design. Parameters are are L$ = 500~\mu$m, $R = 400~\mu$m, $\lambda= 0.866~\mu$m.}
  \label{fig:modes}
\end{figure}

An example for a target mode which contains an equal superposition of $N=3$ LG modes, Eq.\ (\ref{eq:anzats}), is shown in Fig.\ \ref{fig:modes}. Figure (a) shows that the cavity supports exactly one low-loss eigenmode for the chosen parameters while all other modes exhibit large round trip losses. Figure (b) shows the corresponding values of $|T_i|^2$. The lowest loss mode (eigenmode number 1) is an equal superposition of the first three LG modes with weight 1/3 each, confirming that our mirror design in fact makes this target superposition an eigenmode of the cavity. All other modes contain contributions from high LG modes, which explains their large loss seen in figure (a). For all parameters we tested numerically, we always found that the mirror shapes designed by the procedure described in Section \ref{sec:geom} generate the target cavity eigenmodes.

As discussed above, our calculations are performed within the paraxial approximation. For the parameters of Fig.\ \ref{fig:modes} we found that the error in the obtained coefficients $T_i$ does not exceed $5\%$ when we use the mirror profile extracted from the geometrical phase front (as in Fig.\ \ref{fig:phase}) instead of the thin-mirror approximation. This check shows the limitations of the paraxial approximation together with mode mixing theory.


\begin{figure*}[!tb]
    \centering
    {(a) near concentric, N=3 \hspace{5cm} (b) near confocal, N=3}\\
    \includegraphics[width=0.45\textwidth]{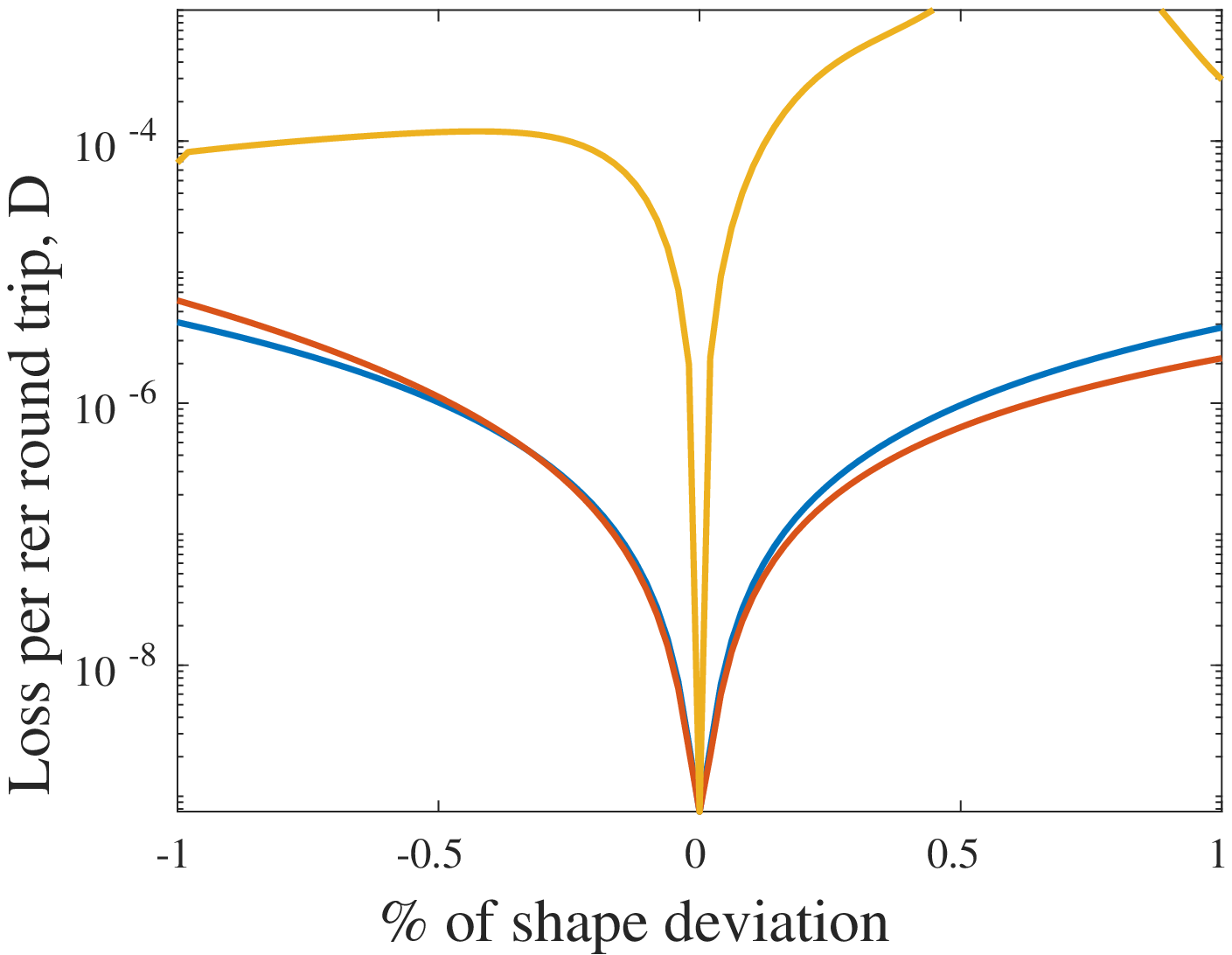}
    \includegraphics[width=0.45\textwidth]{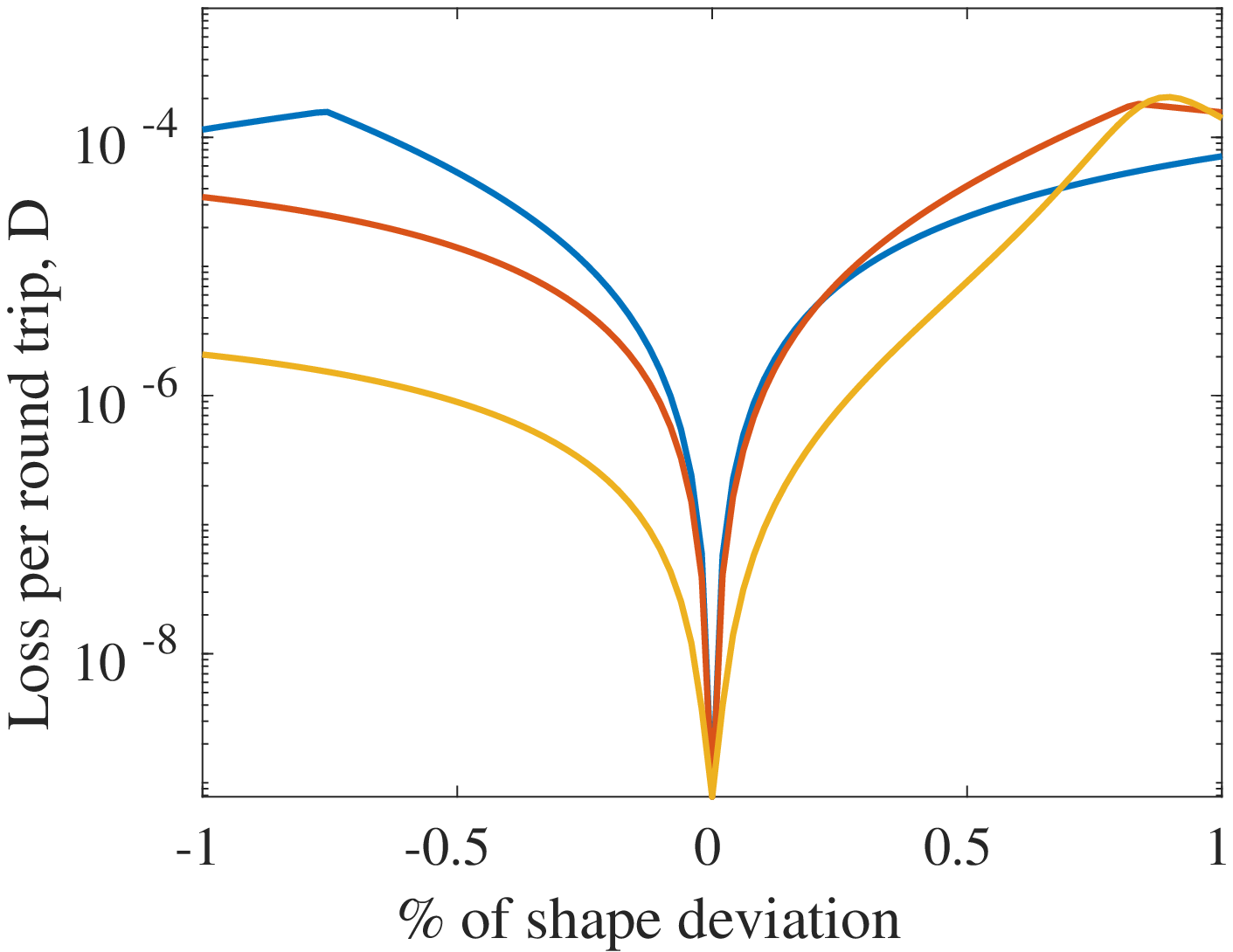}\\
    {(c) near concentric, N=5 \hspace{5cm} (d) near confocal, N=5}\\
    \includegraphics[width=0.45\textwidth]{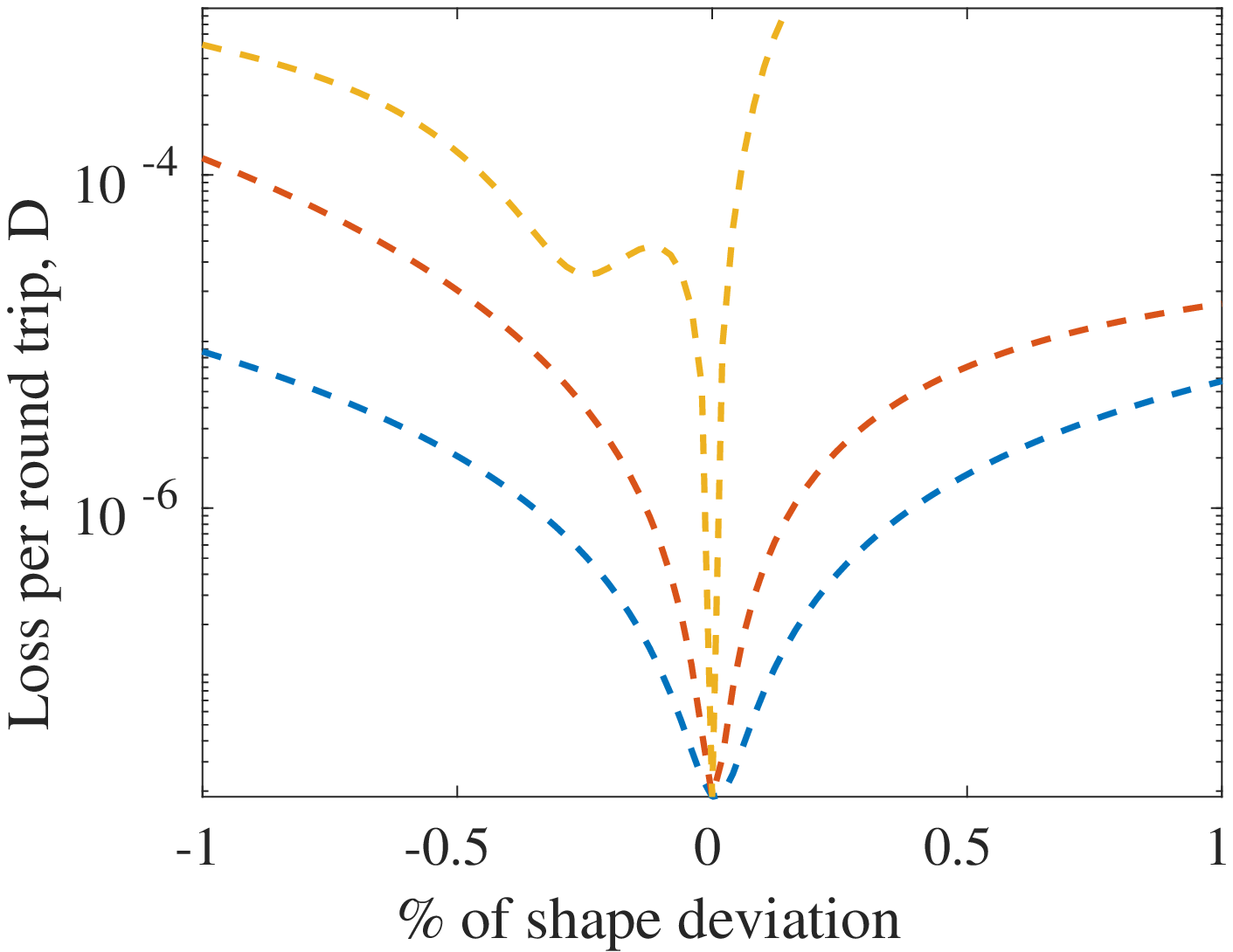}
    \includegraphics[width=0.45\textwidth]{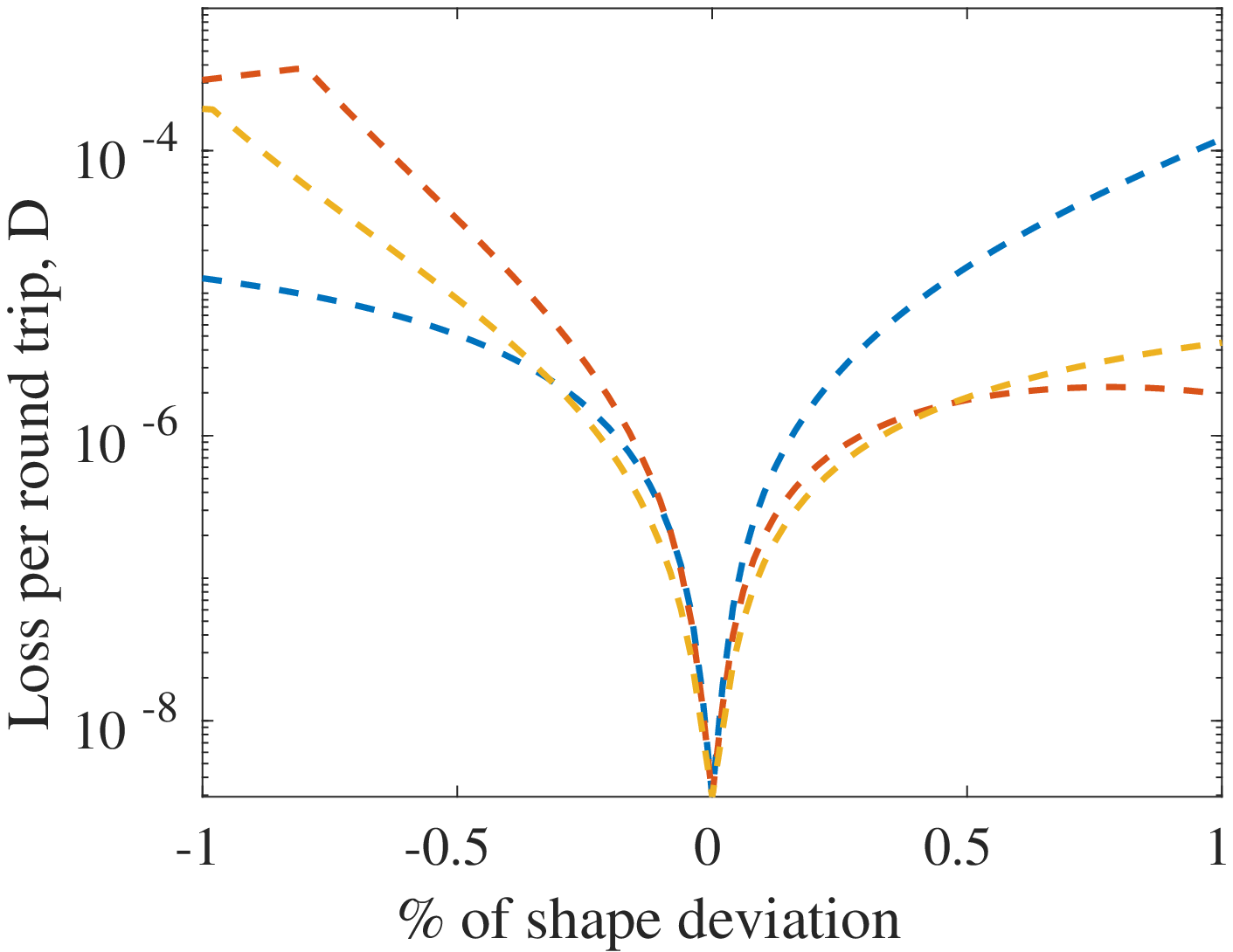}\\
     {(e) near concentric, N=7 \hspace{5cm} (f) near confocal, N=7}\\
    \includegraphics[width=0.45\textwidth]{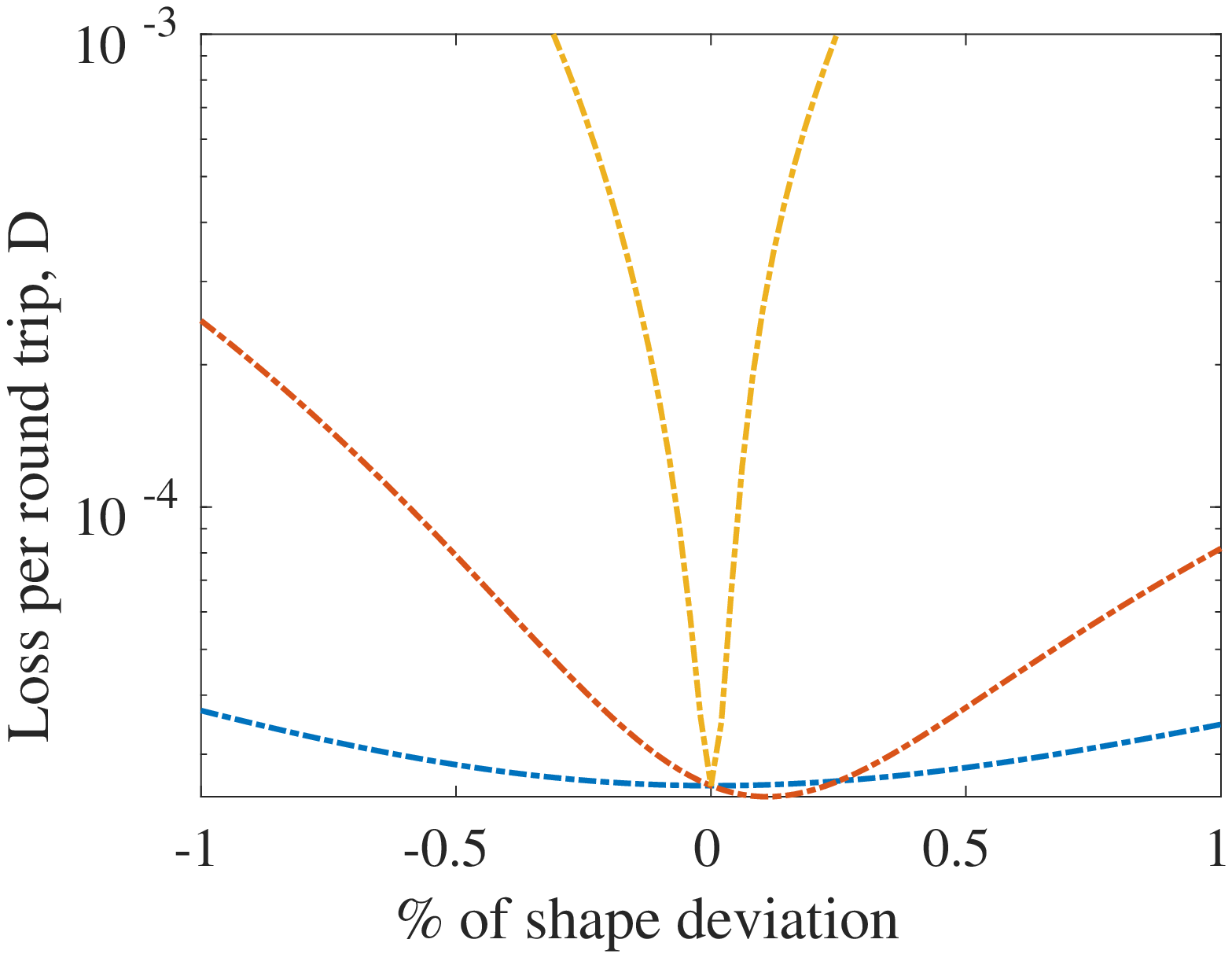}
    \includegraphics[width=0.45\textwidth]{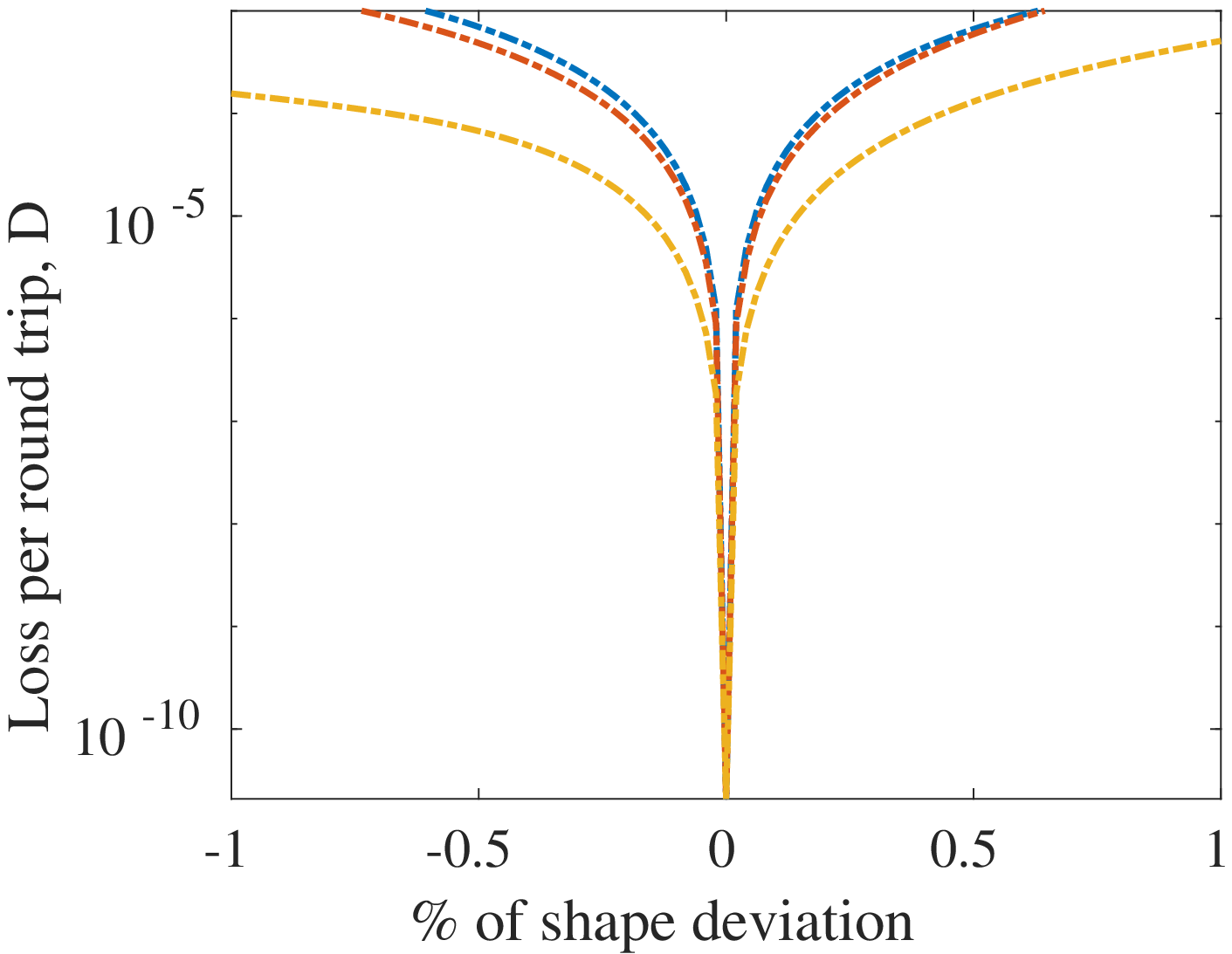}\\
    \caption{Loss per round trip for the lowest loss mode for mirror shapes with systematic fabrication errors. Blue, red and yellow lines correspond to axial, radial and longitudinal misplacements given by deviations of parameters $a$, $b$, and $L$, respectively, see Eq.\ (\ref{eq:deviations}). (a), (c), (e) Near-concentric cavity  with parameters $L = 500~\mu$m, $R = 255~\mu$m. (b), (d), (f) Near-confocal cavity  with parameters $L = 500~\mu$m, $R = 400~\mu$m.  Line styles express different numbers of LG modes in the target eigenmode superposition, solid, dashed and dash-dotted lines correspond to $N = 3, 5, 7$, respectively.}
\label{fig:robust}
\end{figure*}

\subsection{Fabrication Tolerances}\label{robust}

As we discussed in Sec.\ \ref{sec:geom} the target mirror has a spherical shape (parabolic within the paraxial approximation) with some deviations as seen, for example, in Fig.\ \ref{fig:phase}(b). In this section we investigate the effect on cavity eigenmode losses if this target mirror profile is implemented with some errors, e.g., by fabrication imprecision. 

Let us assume that our mirror is described by a profile $P(r)$ as
\begin{equation}
P(r)=-\frac{r^2}{2R} + a \Delta(br),
\label{eq:deviations}
\end{equation}
where the first term is a parabolic mirror with radius of curvature $R$ and the second term is the deviation from the parabolic profile predicted by our theory to generate the desired cavity eigenmode. We assume that the mirrors are fabricated with some systematic error described by two coefficients $a$ and $b$ in Eq.\ (\ref{eq:deviations}) which correspond to axial and radial stretch, respectively, of the target profile $\Delta(r)$ that is machined into the parabolic mirror. We also consider longitudinal misalignment of the cavity, i.e., a cavity length $L$ different from the one used to design the mirrors.

Figure (\ref{fig:robust}) shows the loss of the lowest-loss cavity mode eigenmode when the coefficients $a$ and $b$ in Eq.\ (\ref{eq:deviations}) differ from their design value of 1 and when the cavity length $L$ is slightly varied. Results for two values of the mirror radius of curvature are shown as well as for target mode superpositions containing $N=3,5,7$ LG modes. Overall, we can conclude that manufacturing inaccuracies of the order of $0.5\%$ will not increase the cavity mode loss beyond an intrinsic mirror absorption/transmission loss of $D_{mir} = 10^{- 5}$. Note that according to these simulations an error in cavity length $L$ seems to be the most severe in many cases. However, in practice the cavity length in quantum optics experiments is always controlled to extremely high precision in order to maintain the cavity resonance frequency close to the transition frequency of the quantum particle of the experiment (atom, ion, or quantum dot) and therefore this requirement should not pose any additional restrictions on cavity alignment. 

In general, we find that superpositions of more LG modes are more difficult to achieve with good stability, i.e., they are more susceptible to fabrication errors. However, for the chosen parameters in Fig.\ \ref{fig:robust} it appears interestingly that the near-confocal cavity with a three-mode superposition is less stable than the cavity with a five-mode superposition


\subsection{Comments on Fabrication}\label{fab}

Our scheme requires mirrors that are machined with a small deviation from a spherical (parabolic) profile, as calculated in Section \ref{sec:geom}. We envisage that these profiles are machined into the mirror substrate before the substrate is coated with a high-reflectivity dielectric stack.

Various methods exist that, in principle, allow the fabrication of such fine mechanical structures. Laser micro-fabrication has successfully been demonstrated in silica \cite{laserFab1,laserFab2,laserFab3}. Pulses of a CO$_2$ laser can be used for thermal evaporation of surface material \cite{Hunger2010, KellerFab} and laser radiation focused on the cleaved surface can compose a surface landscape with extremely low roughness. Another method is focused ion beam milling \cite{fib} that can be used for material ablation with nanometer precision. Finally, modern mechanical micro-machining tools \cite{mech1, mech2, mech3} can also provide sufficient precision to meet the fabrication requirements discussed in Section \ref{robust}.


\section{Conclusions}\label{sec:conclusion}

In this work we developed a novel approach that allows us to achieve a significant enhancement of the coupling rate and cooperativity between a quantum emitter and a low-loss optical cavity  without increasing mode instability. We proposed to modify the cavity mirror geometry in such a way that a specific superposition of LG modes becomes an eigenmode of the cavity and provides significant local field enhancement. We demonstrated that such mode superposition can provide lower losses and higher stability at the same electric field amplitude than can be achieved by a more conventional approach using near-concentric cavities. We envisage that such cavities could find widespread use in a variety of research and engineering problems. 

We demonstrated an approach to design the required mirror shapes by following the contours of constant phase of the target mode superposition field and verified our designs using mode matching theory. Finally we briefly discussed options for manufacturing such mirrors and we investigated the effect of systematic errors in fabrication on the performance of the cavity designs. Overall, our approach appears realistic for state-of-the-art manufacturing methods.


\section{Acknowledgments}

We acknowledge financial support by the UK Quantum Technology Programme under the EPSRC Hub in Quantum Computing and Simulation (EP/T001062/1).



\begin{thebibliography}{0}%
\makeatletter
\providecommand \@ifxundefined [1]{%
 \@ifx{#1\undefined}
}%
\providecommand \@ifnum [1]{%
 \ifnum #1\expandafter \@firstoftwo
 \else \expandafter \@secondoftwo
 \fi
}%
\providecommand \@ifx [1]{%
 \ifx #1\expandafter \@firstoftwo
 \else \expandafter \@secondoftwo
 \fi
}%
\providecommand \natexlab [1]{#1}%
\providecommand \enquote  [1]{``#1''}%
\providecommand \bibnamefont  [1]{#1}%
\providecommand \bibfnamefont [1]{#1}%
\providecommand \citenamefont [1]{#1}%
\providecommand \href@noop [0]{\@secondoftwo}%
\providecommand \href [0]{\begingroup \@sanitize@url \@href}%
\providecommand \@href[1]{\@@startlink{#1}\@@href}%
\providecommand \@@href[1]{\endgroup#1\@@endlink}%
\providecommand \@sanitize@url [0]{\catcode `\\12\catcode `\$12\catcode
  `\&12\catcode `\#12\catcode `\^12\catcode `\_12\catcode `\%12\relax}%
\providecommand \@@startlink[1]{}%
\providecommand \@@endlink[0]{}%
\providecommand \url  [0]{\begingroup\@sanitize@url \@url }%
\providecommand \@url [1]{\endgroup\@href {#1}{\urlprefix }}%
\providecommand \urlprefix  [0]{URL }%
\providecommand \Eprint [0]{\href }%
\providecommand \doibase [0]{https://doi.org/}%
\providecommand \selectlanguage [0]{\@gobble}%
\providecommand \bibinfo  [0]{\@secondoftwo}%
\providecommand \bibfield  [0]{\@secondoftwo}%
\providecommand \translation [1]{[#1]}%
\providecommand \BibitemOpen [0]{}%
\providecommand \bibitemStop [0]{}%
\providecommand \bibitemNoStop [0]{.\EOS\space}%
\providecommand \EOS [0]{\spacefactor3000\relax}%
\providecommand \BibitemShut  [1]{\csname bibitem#1\endcsname}%
\let\auto@bib@innerbib\@empty
\end{thebibliography}%


\begin{thebibliography}{10}

\bibitem{Pellizzari1995}
Pellizzari T, Gardiner S, Cirac J and Zoller P 1995 {\it Phys. Rev.
  Lett.\/} {\bf 75} 3788--3791

\bibitem{Cirac1997}
Cirac J~I, Zoller P, Kimble H~J and Mabuchi H 1997 {\it Phys. Rev.
  Lett.\/} {\bf 78} 3221--3224

\bibitem{Kimble2008}
Kimble H~J 2008 {\it Nature\/} {\bf 453} 1023--1030

\bibitem{Monroe2013}
Monroe C and Kim J 2013 {\it Science\/} {\bf 339} 1164--1169

\bibitem{fib}
Romagnoli, P., Maeda, M., Ward, J.M. et al. Fabrication of optical nanofibre-based cavities using focussed ion-beam milling: a review. Appl. Phys. B 126, 111 (2020).

\bibitem{revResonators}
Bitarafan MH, DeCorby RG. On-Chip High-Finesse Fabry-Perot Microcavities for Optical Sensing and Quantum Information. Sensors (Basel). 2017 Jul 31;17(8):1748.


\bibitem{cooling}
Laser cooling and trapping of neutral atoms Nobel Lecture by William D. Phillips, Dec 8, 1997: Phillips, William D. (1998). "Nobel Lecture: Laser cooling and trapping of neutral atoms". Reviews of Modern Physics. 70 (3): 721–741. Bibcode:1998RvMP...70..721P. doi:10.1103/RevModPhys.70.721.

\bibitem{prep}
I. H. Deutsch and P. S. Jessen, "Quantum state preparation in optical lattices," in Quantum Electronics and Laser Science Conference, J. Bokor, R. Slusher, P. Bucksbaum, and R. Falcone, eds., Vol. 12 of OSA Technical Digest (Optical Society of America, 1997), paper QTuJ3.

\bibitem{comp}
D. M. Lucas, C. J. S. Donald, J. P. Home, M. J. McDonnell, A. Ramos, D. N. Stacey, J.-P. Stacey, A. M. Steane, S. C. Webster
Philos. Transact. A Math Phys Eng Sci. 361, 1401-8

\bibitem{Harlander2010}
Harlander M, Brownnutt M, H{\"{a}}nsel W, R and Blatt R 2010 {\it New J.\
  Phys.\/} {\bf 12} 093035

\bibitem{Podoliak2016}
Podoliak N, Takahashi H, Keller M and Horak P 2016 {\it Phys.\ Rev. Appl.\/} {\bf 6} 044008
  
  
\bibitem{Vucovich}
Jelena Vučković, Stanford University
Lectures given at Les Houches 101th summer school on “Quantum Optics and Nanophotonics", August 2013 (to be published by Oxford University Press)  

\bibitem{yariv}
Amnon Yariv, Quantum Electronics, 3rd Edition, ISBN: 978-0-471-60997-1, January 1991, 704 Pages


\bibitem{Kleckner2010}
Kleckner D, Irvine W~T~M, Oemrawsingh S~S~R and Bouwmeester D 2010 {\it Phys.\
  Rev.\ A\/} {\bf 81} 043814
  
\bibitem{Nina}
Nina Podoliak, Hiroki Takahashi, Matthias Keller and Peter Horak, J. Phys. B. 50, 085503 (2017)

\bibitem{Benedikter2015}
Benedikter J, H\"ummer T, Mader M, Schlederer B, Reichel J, H\"ansch T~W and
  Hunger D 2015 {\it New J.\ Phys.\/} {\bf 17} 053051

\bibitem{CaIon}
B. Brandstätter, A. McClung, K. Schüppert, B. Casabone, K. Friebe, A. Stute, P. O. Schmidt, C. Deutsch, J. Reichel, R. Blatt, and T. E. Northup , "Integrated fiber-mirror ion trap for strong ion-cavity coupling", Review of Scientific Instruments 84, 123104 (2013) 

\bibitem{Phillips1983}
Phillips, R.L., and Andrews, L.C. Spot size and divergence for Laguerre Gaussian beams of any order. Appl. Opt. 22, 643-644 (1983) 

\bibitem{Lax1975}
From Maxwell to paraxial wave optics
Melvin Lax, William H. Louisell, and William B. McKnight
{\it Phys. Rev. A} 11, 1365 (1975)


\bibitem{laserFab1}
Hunger D, Deutsch C, Warburton R and Reichel J, Laser micro-fabrication of concave, low-roughness features in silica, AIP Advances 2, 012119 (2012)

\bibitem{laserFab2}
Vernooy D W, Furusawa A, Georgiades N P, Ilchenko V S and Kimble H J 1998 Phys. Rev. A 57 R2293

\bibitem{laserFab3}
Armani D K, Kippenberg T J, Spillane S M and Vahala K J 2003 Nature 421 925

\bibitem{Hunger2010}
Hunger D, Steinmetz T, Colombe Y, Deutsch C, H{\"{a}}nsch T~W and Reichel J
  2010 {\it New J.\ Phys.\/} {\bf 12} 065038

\bibitem{KellerFab}
Gulati, G.K., Takahashi, H., Podoliak, N. et al. Fiber cavities with integrated mode matching optics. Sci Rep 7, 5556 (2017). https://doi.org/10.1038/s41598-017-05729-8


\bibitem{mech1}
Brinksmeier Ekkard and Preuss Werner, Micro-machining, Phil. Trans. R. Soc. A., 3703973–3992 (2012).

\bibitem{mech2}
Gao, S., Huang, H., Recent advances in micro- and nano-machining technologies., Front. Mech. Eng. 12, 18–32 (2017).

\bibitem{mech3}
Schneider, F., Das, J., Kirsch, B. et al., Sustainability in Ultra Precision and Micro Machining: A Review., Int. J. of Precis. Eng. and Manuf.-Green Tech., 6 601–610 (2019).

\end{thebibliography}
\end{document}